\documentclass[aps,prl,preprint,groupedaddress,showkeys,showpacs]{revtex4}

\begin{document}

\title{The Fermi-Pasta-Ulam problem: 50 years of progress}


\author{G.~P.~Berman and F.~M.~Izrailev\thanks{1}}
\thanks{on sabbatical leave from: Instituto de
F\'{\i}sica, Universidad Aut\'{o}noma de Puebla, Apartado Postal
J-48, Puebla, Pue., 72570, M\'{e}xico}
\affiliation{Theoretical Division and CNLS, Los Alamos National
Laboratory, Los Alamos, New Mexico 87545, USA}


\date{\today}

\begin{abstract}
A brief review of the Fermi-Pasta-Ulam (FPU) paradox is given,
together with its suggested resolutions and its relation to other
physical problems. We focus on the ideas and concepts that have
become the core of modern nonlinear mechanics, in their historical
perspective. Starting from the first numerical results of FPU, both
theoretical and numerical findings are discussed in close connection
with the problems of ergodicity, integrability, chaos and stability
of motion. New directions related to the Bose-Einstein condensation
and quantum systems of interacting Bose-particles are also
considered.

\end{abstract}

\pacs{05.45.-a; 63.20.Ry; 95.10.Fh; 03.75.Fi}
\keywords{Fermi-Pasta-Ulam, chaotic dynamics, solitons, energy
equipartition, dynamical chaos}

\maketitle

\section{I. Introduction}

The goal of this paper is two-fold. First, we evaluate and summarize
the most interesting results related to the FPU model, after the
seminal paper \cite{FPU55} appeared in 1955. Second, we discuss new
directions in the study of many-body chaos, that are related,
directly or indirectly, to the FPU problem. We hope that our
analysis will help future investigations of nonlinear classical and
quantum systems of interacting particles.

Numerous attempts to resolve the FPU paradox have resulted in a
burst of analytical and numerical studies of nonlinear effects in
physical systems. The primary interest of FPU was the observation of
energy sharing in one-dimensional lattices with nonlinear coupling
among rigid masses. For Fermi this study was directly related to one
his first papers of 1923 \cite{F23} in which he tried to rigorously
prove the ergodicity hypothesis which lies at the core of
traditional statistical mechanics. For a long time, the ergodicity
was assumed to serve as the only mechanism needed for the foundation
of statistical mechanics. Specifically, by assuming the ergodic
motion of classical trajectories on the surface of constant energy,
one can expect statistical behavior of a system and apply well
developed statistical methods.

In view of the result of Fermi \cite{F23}, and in accordance with
wide-spread expectations, any weak nonlinear interaction between
particles in a system with very many degrees of freedom causes
ergodic behavior of a system. In fact, this point is used in the
derivation of statistical distributions in the thermodynamic
$N\rightarrow \infty$ limit for systems of noninteracting particles.
Therefore, it is natural to expect that a system of 32 or 64
particles, as in the FPU numerical study, would reveal ergodic
behavior, provided the nonlinearity is not extremely small. To the
great surprise of FPU, the result was opposite: the long-time
dynamics of the studied model appeared to be periodic, with almost
perfect returns to the initial conditions. Having extraordinary
intuition, Fermi noted that this effect may have very important
consequences \cite{T68}. About ten years later, two alternative
explanations of the FPU paradox were suggested, giving rise to new
phenomena, the integrability of nonlinear equations and dynamical
(deterministic) chaos.

One of the discoveries triggered by the FPU results was the complete
integrability of a class of nonlinear differential equations. The
first indication of this unexpected fact was due to numerical
integration of the Korteweg-de-Vries equation in 1965 by Zabusky and
Kruskal \cite{ZK65}. As was demonstrated, stable solitary waves
({\it solitons}) emerged from generic initial conditions and
traveled through the media, interacting with each other without
losing their identities. This effect was suggested as an explanation
of the remarkable recurrence in the FPU model, with the claim of its
closeness to a completely integrable model.

Another approach in resolving the FPU paradox was developed by
Chirikov on the basis of his criterion of {\it stochasticity} (or
{\it dynamical chaos}) \cite{C59}. As was already known from the
study of nonlinear systems in applications to accelerator and plasma
physics, the motion of a dynamical nonlinear system with few degrees
of freedom can exhibit strong chaotic behavior.  Thus, the
description of these systems can be given in terms of conventional
statistical mechanics. The mechanism of chaotic behavior of
dynamical systems was found to be an exponential instability of
motion for a wide range of initial conditions. The essential role in
the emergence of this kind of instability is played by interacting
nonlinear resonances. By 1965, the relatively simple criterion of
resonance overlap enabled one to determine the conditions for the
onset of stochasticity for various low-dimensional systems.
Therefore, it was natural to apply the same approach to nonlinear
systems of the FPU type. As a result, the threshold of stochasticity
was found analytically in Ref.\cite{IC65} and later confirmed
numerically \cite{IKC68}. According to these studies, the initial
conditions used by FPU in their numerical simulations were chosen
below the stochasticity threshold, just in the region corresponding
to stable quasi-periodic motion. Above this threshold, the FPU model
was shown to behave in accordance to the original expectations of
FPU, revealing strong statistical properties, such as energy
equipartition among the linear modes.

In fact, the FPU study gave birth to a new method to study the
physical laws of nature. The two first methods are well-known:
theoretical and experimental physics. The new approach was predicted
by Ulam and discussed in his mathematical book \cite{U60}. Expecting
a future burst of computer technology, he proposed a new kind of
{\it synergetic} cooperation between a physicist and a computer
(see, also, Ref.\cite{Z81}). Apart from the normal use of computers
as a tool for the calculations of integrals, functions, differential
equations, etc., the new role for computers consists of the study of
physical systems {\it ab initio}, starting from given models and
investigating their properties by varying parameters, forces,
initial conditions, etc. This kind of activity was marked in the FPU
paper as ``numerical experiments." Later, this term has been widely
used by Chirikov to stress the difference of the new approach from
both theoretical and experimental studies.

The FPU problem may be treated as a perfect example of this
approach. Specifically, first, the model was set up in the form of
equations of motion. Then, ``numerical experiments" were performed
to see how rapidly the thermalization occurs due to the nonlinear
interactions. And, unexpectedly, a new phenomenon was discovered,
initiating further theoretical studies. Afterwards, theoretical
predictions were checked and further numerical studies gave new
insight in the problem. Thus, the ``synergetic" approach progresses.
Since the time of the first ``numerical experiments" of FPU many
physical discoveries have been found first numerically, then
explained theoretically and confirmed by real experiments. An
exciting story of first twenty years of studies of the FPU paradox
can be found in the book of Weissert \cite{W97}.

\section{II. The FPU model}

The primary aim of the authors of Ref.\cite{FPU55} was to observe
thermal equilibrium in a nonlinear string, and to establish the rate
of approach to the equipartition of energy among different degrees
of freedom. In order to treat this problem numerically, the
continuum was represented by a large number of masses interacting
with each other via nonlinear forces.

The corresponding partial differential equations were approximated
by a linear chain of particles of equal masses $M$ connected by
elastic springs. The linear part of the forces is determined by the
constant $K$, resulting in harmonic frequencies
$\omega_0=\sqrt{2K/M}$ for all particles. (Following many papers we
assume, for simplicity, $M=K=1$). For the nonlinear part, in
Ref.\cite{FPU55} main attention was paid to the simplest cases of
quadratic and cubic additional terms, although some alternative
forms of the interaction have also been discussed. For the quadratic
force (called the $\alpha-${\it model}) the corresponding equations
of motion are
\begin{equation}
\label{alpha}
\ddot{x}_n=(x_{n+1}-2x_n+x_{n-1})+\alpha[(x_{n+1}-x_n)^2-(x_n-x_{n-1})^2].
\end{equation}
Correspondingly, the chain of particles with additional cubic forces
(called the $\beta-${\it model}) is governed by the equations
\begin{equation}
\label{beta}
\ddot{x}_n=(x_{n+1}-2x_n+x_{n-1})+\beta[(x_{n+1}-x_n)^3-(x_n-x_{n-1})^3]
\end{equation}
Here $x_n$ denotes the displacement of the $n-$th particle from its
original position, and the parameters $\alpha$ and $\beta$ are the
strengths of nonlinear interactions between particles.

In the absence of nonlinear forces the exact solution can be written
in the form of normal modes $Q_k(t)$ that are essentially the
Fourier representation of the displacements $x_n(t)$ (for fixed ends
of the chain, $x_0=x_N=0$),
\begin{equation}
\label{modes} Q_k(t)=\sqrt{\frac{2}{N}}\sum_{n=1}^N x_n(t)
\sin{\frac{\pi kn}{N}}
\end{equation}
With this representation one can see that for any initial conditions
$x_n(0)$ and $\dot{x}_n(0)$ the energy
\begin{equation}
\label{Emode} E_k=\frac{1}{2}(\dot {Q}_k^2+\omega_k^2Q_k^2)
\end{equation}
of every $k-$th mode is constant. Therefore, the model is trivially
integrable and energy equipartition among normal modes is not
possible. As a result, the motion of this model is quasi-periodic in
time, with the discrete spectrum determined by the normal
frequencies $\omega_k$,
\begin{equation}
\label{normal_omega} \omega_k=2\sin{\left(\frac{\pi k}{2N} \right)
}.
\end{equation}
In the normal mode representation, the equations of motion take the
form
\begin{equation}
\label{alphaq} \ddot{Q}_k+\omega_k^2 Q_k= \alpha \sum_{i,j=1}^N
C_{ij} Q_i Q_j
\end{equation}
for the $\alpha-$model (\ref{alpha}) and
\begin{equation}
\label{betaq} \ddot{Q}_k+\omega_k^2 Q_k= \beta \sum_{i,j,l=1}^N
D_{ijl} Q_i Q_j Q_l
\end{equation}
for the $\beta-$model (\ref{beta}). Here $C_{i,j}$ and $D_{i,j,l}$
are coefficients of the complicated dependence on the indexes $i,j$
and $l$, defined by the nonlinear forces.

Having in mind the predictions of conventional statistical
mechanics, the authors of Ref.\cite{FPU55} expected that by
switching on the nonlinear terms in Eqs.(\ref{alpha})-(\ref{beta})
[or equivalently Eqs.(\ref{alphaq})-(\ref{betaq})], energy initially
concentrated in a particular mode, will flow into all other modes,
thus demonstrating the transition to equilibrium. In particular,
analytical arguments were given in Ref.\cite{P62}, according to
which after a long time the systems of coupled anharmonic
oscillators have to approach thermal equilibrium. Thus, it was a
general belief that any kind of nonlinearity in a system with large
number of degrees of freedom would give rise to ergodicity (see,
e.g., Ref.\cite{F23}). And the latter was assumed to serve as the
mechanism for the onset of statistical behavior in dynamical
systems.

The numerical studies in Ref.\cite{FPU55} were performed on
Metropolis' new MANIAC computer with $N=32$ or $N=64$ and with
sufficiently small values of the nonlinear parameters $\alpha$ and
$\beta$, for zero initial conditions, $x_0=x_N=0$. The first results
refer to 1953-1954, with some additional runs that have been done
after the death of Fermi in December 1954. Typically, the first mode
with $k=1$ was initially excited, for which most details are given.
The time dependence of the energies $E_k(t)$ of all modes was
studied for many fundamental periods $T_1=2\pi/\omega_1$ (for more
details, see, e.g. Ref.\cite{F92}).

To the great surprise of the authors of Ref.\cite{FPU55}, the
results of a numerical simulation were quite astonishing. The
behavior of both models was at first as expected: the energy spread
to higher harmonics but after about 1000 oscillation periods $T_1$,
the flow of the energy into other modes stops, and the dynamics
reversed, with the energy flowing back into the first mode. This
recurrence of energy was found to be almost complete, with a
decrease in energy of only about $2\%$ of the total energy. In time,
this periodic behavior persisted, thus demonstrating the absence of
the expected statistical thermalization. The surprise was enhanced
by the fact that the period of a recurrence was found to decrease
with increasing coefficients of nonlinearity. Therefore, the
nonlinear effects are significant and cannot be neglected. The time
evolution did not lead to the equipartition of energy, rather, it
demonstrated the existence of ``quasi-modes" consisting of a number
of linear modes. According to Tuck (\cite{T68}), Fermi became really
excited about this phenomena and thought that ``something new and
important might be at hand."

A few possible reasons for this observed effect were discussed
during the first stage of the story. Initially, the accuracy of the
numerics was questioned, with a hint that more accurate calculations
would show thermalization, although a very weak one. This point was
somehow supported by the observation of a non-complete return of the
energy in the originally excited mode. However, further more
accurate computations of Tuck in 1961 (see Refs.\cite{T68,TM72})
revealed an even more exciting effect. It was found that at later
times, the recurrence of the energy becomes more nearly complete.
Specifically, a ``super period" was found that is about 80 000
linear cycles, $T_1$. The energy recurrence after this super period
was more than 99\% of the total energy. In general, these results of
Tuck, although not discussed in the literature until much later
\cite{TM72}, confirmed the phenomena of the recurrence in the
FPU-model.

Of special interest was whether or not the energy recurrence can be
associated with Poincar\'e cycles that occur for ergodic systems.
The estimate of the Poincar\'e cycle for a chain of linear
oscillators was derived in Ref.\cite{HMW58}. This estimate shows
that the recurrence time in a chain of linear oscillators increases
in an approximately exponential way with the number of degrees of
freedom. Therefore, it is clear that the Poincar\'{e} cycles have no
relation to the observed recurrence in the FPU-model. A careful
analysis \cite{J78} of this estimate in application to the
FPU-problem shows, however, that one should distinguish the
FPU-recurrence from the Poincar\'e cycles. The point is that the
latter are defined for trajectories in phase space, rather than for
the energy of a system. Obviously, the energy recurrence time will
usually be much less than the Poincar\'e time. Moreover, the
estimate was given for a harmonic lattice, and there is no way,
apart from direct computation, to determine this recurrence time in
the presence of nonlinear coupling. This remark, however, does not
change the conclusion that the FPU-recurrence has a different nature
than the Poincar\'e cycles.

In view of the many discussions about the mechanism of
irreversibility in the systems of interacting particles, it is
instructive to mention some of the computations of Tuck. To see the
influence of numerical errors, he performed the following check.
After a few thousand cycles, the dynamics of the model was
numerically reversed by the change of time and velocities of all
particles. It was then found that 100\% of the energy returns to the
first mode. This fact was underestimated in the early 1960s. Now it
can be treated as a direct (numerical) proof of the {\it regular}
dynamics in the above models.

As is now well known, dynamical chaos is characterized by an
exponential sensitivity of the dynamics to the initial conditions.
As a result, the unavoidable round-off errors in numerical
simulations give rise to a drastic change for individual chaotic
trajectories. Because of this exponential sensitivity and the
round-off errors it is not possible to reach numerically the initial
state, unless the reversal time is very small. This fact leads to
the very important conclusion that chaotic systems cannot be treated
as isolated ones since any weak external perturbation is essentially
strong (see the discussion in Refs.\cite{C69,C79}). Therefore, this
local instability serves as a mechanism for the apparent
irreversibility in dynamical systems, although any dynamical system
is reversible in principal (here, we do not discuss dissipative or
noisy systems).

For about a decade after the publication of the FPU preprint,
discussions of the FPU paradox were restricted to a trivialization
of the results, attempting to explain the recurrence effect as
simply as due to numerical errors, insufficient computation time,
Poincar\'e recurrence or the specific choice of nonlinear forces
which prevents the ergodicity. It was still not well recognized that
the FPU results initiated a new era in physics, associated with
nonlinear phenomena.

\section{III. Perturbative approaches}

The first analytical studies of the FPU paradox were described in
the paper of Ford \cite{F61}. It was argued there that the absence
of ergodicity in the FPU calculations may be due to the arithmetic
properties of the unperturbed spectrum determined by
Eq.(\ref{normal_omega}). By making use of the perturbation theory of
Kryloff and Bogol\`iuboff \cite{KB47}, it was claimed that
appreciable energy sharing among normal modes for a very weak
coupling nonlinear interaction occurs only if the frequencies
$\omega_k$ of the unperturbed motion are linearly dependent (or,
only if $\sum_k m_k \omega_k = 0$ for some nonzero collection of
integers $\{m_k=0\})$. As for the FPU numerical data, they refer to
the value of $N$ as a power $2$, therefore, to linearly independent
frequencies (see details in Ref.\cite{F61}). For this reason, only
few (low) modes in the FPU simulation could share the energy.
Therefore, one should have multiple resonance conditions, in order
to expect widespread energy sharing. However, as was shown in
Ref.\cite{J63b}, this idea, although quite useful in the description
of weakly nonlinear oscillations, turned out to fail to explain the
FPU paradox. Numerical experiments with many other values of $N$
confirmed irrelevance of linear resonance conditions to the FPU
recurrence.

The numerical data of Ref.\cite{FPU55} describe a relatively weak
interaction between particles. This fact has triggered analytical
studies of the FPU dynamics utilizing perturbation theories. In an
attempt to explain the quasi-period for the normal modes, in
Ref.\cite{J63a} standard perturbation methods were examined in light
of the application to long-time dynamics of nonlinearly coupled
oscillators. As is known, the main problem is the small divisors
that arise due to resonances between unperturbed oscillators. In the
large $N-$limit the frequencies become dense and the frequency
differences approach zero, causing all terms containing the small
divisors to become infinite. Another problem is related to the
appearance of secular terms that are proportional to a power of time
$t$, and, therefore, restrict the application of time-dependent
expressions to finite times. In order to avoid these secular terms,
the Kryloff-Bogoliubov method \cite{KB47} was modified \cite{J63a}
and applied \cite{J63b} to the FPU model. Specifically, second-order
perturbation theory was found to give an accurate estimate (within
15\%) of the recurrence time and amount of energy exchange in the
FPU problem. On the other hand, it was revealed that for some cases
numerically studied in Ref.\cite{FPU55}, a higher-order analysis is
required due to a relatively large nonlinearity (when the nonlinear
term is of the order of one-tenth of the linear term, in energy
units). The important conclusion of these studies is that, strictly
speaking, the FPU model does not belong to the category of weak
coupling.

It was also indicated \cite{J63b} that when discussing the limit $N
\rightarrow \infty$, one should distinguish two different limits.
The first one considered in Ref.\cite{Z62} (see the discussion
following), assumes that the length $L=Na$ of the chain remains
constant due to the decreasing spacings, $a$, between the particles.
Correspondingly, the effective coupling $\bar\alpha$ decreases with
$N$ as $2\alpha/N$ (correspondingly, $\bar\beta=3\beta/N^2$). In
this way, by normalizing time $t \rightarrow tN$ and the spacial
coordinate $z \rightarrow zL^{-1}$, one can obtain the following
partial differential equations:
\begin{equation}
\label{alphaC} \frac {\partial^2 x} { \partial t^2} =  \left[ 1 +
\bar\alpha \frac {\partial x} { \partial z} \right] \frac
{\partial^2 x} {\partial z^2}
\end{equation}
and
\begin{equation}
\label{betaC} \frac{ \partial^2 x} { \partial t^2} =  \left[ 1 +
\bar\beta \left( \frac {\partial x} {\partial z} \right)^2 \right]
\frac {\partial^2 x} {\partial z^2}.
\end{equation}
The corresponding initial conditions are prescribed over the range
$0 < z < 1 $ as $x(z,0) = x_0(z)$ and $\partial x / \partial t
|_{t=0} = 0$.

The other possible limit assumes the parameters of the chain are
constant. Therefore, as $N \rightarrow \infty$, the length $L$
becomes infinite and the frequency spectrum becomes dense. This is
the limit typically discussed in the literature, especially, for the
study of irreversible processes. One of the main questions is how
the  statistical properties of this system depends on the number $N$
of interacting particles.

As is shown in Ref.\cite{J63b}, the phenomenon of recurrence in
nonlinearly coupled oscillators is quite robust. Namely, asssuming
the coupling constant in Eq.(\ref{alpha}) to be different for
different particles (a kind of imperfection), one can ``kill" the
recurrence only with a sufficiently strong imperfection. This
important fact indicates that the FPU recurrence is not an artifact
of the chosen forces between particles.

The meaning of the FPU recurrence and its relevance to the problem
of ergodicity was thoroughly discussed in Ref.\cite{FW63}. Taking
the results of FPU to be fundamental, it was suggested that
ergodicity may not be required for the onset of thermalization in
dynamical systems. In fact, this was a new insight into the problem
of the foundations of statistical mechanics. In support of this
point, it was noted that even a completely integrable system of
linearly coupled oscillators shows a kind of thermalization for
generic initial conditions. Specifically, after initially exciting
one particular mass in the linear chain, one can observe an
effective energy sharing among all particles. Having this analogy in
mind, the authors of Ref.\cite{FW63} performed an analytical study
of energy sharing between linear modes in the $\alpha-$model
(\ref{alpha}) with 2, 3, 5 and 15 oscillators. One of the results of
this analysis was a modification of the resonance condition obtained
previously in Ref.\cite{F61} required for strong energy sharing,
$\sum m_k\omega_k \lesssim \alpha$, where $m_k$ are nonzero integers
which depend on the particular coupling used. It was also found
that, apart from this condition, strong energy sharing occurs for
only certain initial conditions, not for all conditions. As a
result, it was concluded that the dynamics of the FPU model may be
consistent with the existence of additional integrals of motion;
however, one can still speak about thermalization. To check this,
the distribution of linear mode energies $E_k=\dot Q^2+\omega_k^2
Q^2$ was obtained numerically for $n=5$ by examining the time
dependence of $E_k(t)$. A very good correspondence to the
exponential dependence was found, in accordance with the predictions
of statistical mechanics.

Thus, a new approach to thermal equilibrium problem was suggested:
instead of the search of the ergodicity, one should study the
conditions for strong energy sharing between normal modes. Moreover,
it was suggested that in addition to the total energy, other
integrals of motion may exist. At least one integral of motion was
indicated to exist \cite{FW63}, due to a peculiarity of the model
(the unperturbed part is purely linear, unlike many other examples
for which in the absence of perturbation the unperturbed motion is
nonlinear, see the discussion following).

A new viewpoint according to which typical nonlinear systems are
non-ergodic, has found rigorous confirmation based on the extensive
mathematical studies of Kolmogorov, Arnold and Moser (KAM theory,
\cite{K54,A63,M62}, see, also, in Ref.\cite{AA68}). In 1954
Kolmogorov formulated the theorem that states that a weak nonlinear
perturbation of an integrable system destroys the constants of
motion only locally in the regions of resonances. In other regions
of phase space, a set of points of positive measure remains for
which quasi-periodic motion persists. This effect occurs for quite
generic conditions on both the unperturbed motion and the type of
perturbation. Loosely speaking, these conditions are as follows: the
unperturbed system has to be nonlinear and the perturbation has to
be weak enough and with a sufficient number of continuous
derivatives. In the first proof of Arnold of Kolmogorov's theorem
\cite{A63}, for technical reasons, the number $M$ of derivatives was
assumed to be vary large. However, in subsequent studies by Moser
\cite{M62}, the minimal value of $M$ was significantly reduced (see,
also, Refs.\cite{C69,C79} where corresponding estimates of this
number were discussed and improved). Although a direct application
of KAM theory to the FPU model is questionable (see the discussion
following), the main result according to which one should not expect
the ergodicity for a weak nonlinear perturbation, was essential for
the acceptance of a non-ergodic dynamics in nonlinear lattices.

Another perturbative approach has been suggested in Ref.\cite{N71}.
It is based on the concept of Birkhoff-Gustavson normal forms as
approximations to the Hamiltonians of the FPU lattices. Using these
forms, one can show that for weak nonlinearity the motion of the FPU
model is near-integrable. Further developments of this approach are
reported in Refs.\cite{RV00,R01} where the role of discrete
symmetries and resonances was examined in great detail. In was also
claimed that with the use of normal forms, the KAM theorem can be
verified.

In view of the KAM theory, it is important to mention the paper
\cite{NP64} in which energy sharing and equilibrium were numerically
studied for a chain of particles with elastic collisions.
Specifically, in addition to linear forces between particles,
elastic collisions were assumed caused by the finite diameters of
particles. In contrast to the recurrence dynamics in the FPU model,
in Ref.\cite{NP64} behavior that can reasonably be described as
ergodic was found for $N=3$ up to $N=32$ particles. This ergodicity
was ovserved both in the equipartition of the energy among all
linear modes, in the time average, and by a rapid relaxation to
equilibrium with the predicted values of temperature and pressure.
This result was extremely important for the understanding that the
type of interaction between particles plays an essential role. As
was understood later, the numerical data of Ref.\cite{NP64} are in
agreement with a rigorous proof by Sinai \cite{S66} of ergodicity
for a system of hard spheres with elastic collisions contained in a
box. Moreover, apart from ergodicity, {\it mixing} was proved in
Ref.\cite{S66} which currently  is considered as the most important
property of dynamical chaos. Note that mixing automatically implies
ergodicity.

\section{IV. Integrability}

Inspired by the FPU paradox, in 1962 Zabusky analytically studied
the continuous limit (\ref{alphaC}) of the $\alpha-$model
(\ref{alpha}). He found an exact analytical solution for fixed
initial conditions which turns out to break down at time $t_c \sim
1/\bar \epsilon$. At this time, the first derivative $x_z$ develops
a discontinuity, therefore, $x_{zz}$ becomes singular. This means
that the long-time dynamics of the FPU cannot be approximated by the
differential equations (\ref{alphaC}). On the other hand, for time
scales less than the breakdown time $t_c$ the comparison of the
analytical solution with numerical data of Ref.\cite{FPU55} was
reasonable. As was found in \cite{Z62,KZ64}, this critical time
$t_c$ corresponds approximately to the time at which the energy in
the second mode reaches its first maximum (if only the mode with
$k=1$ is initially excited).

Further analysis \cite{KZ64} showed that to study analytically what
happens in the related continuous model, one must include the higher
spatial derivatives that were omitted in taking the lowest continuum
limit (\ref{alphaC})-(\ref{betaC}). To do this, one should use the
following Taylor expansion of spatial and temporal differences:
\begin{equation}
\label{expantion} x_{n\pm1} -x_n = \left [ \pm a x_z + \frac
{a^2}{2} x_{zz} \pm \frac { a^3}{6} x_{zzz} + \frac{a^4}{24}
x_{zzz}. . . \right ]_{z=z_n}
\end{equation}
where $a=L/N$ and $x(z)=x_n, \, z=na$. Therefore, the corresponding
equation for the $beta-$model takes the form
\begin{equation}
\label{zab} \frac{ \partial^2 x} { \partial t^2} =  \left[ 1 +
\bar\beta \left( \frac {\partial x} {\partial z} \right)^2 \right]
\frac {\partial^2 x} {\partial z^2} + \frac {a^2}{12} \frac{
\partial^4 x} {\partial x^4}
\end{equation}
which descibes shallow water waves in classical hydrodynamics (see
the discussion in Ref.\cite{T75}).

For traveling waves in one direction only (e.g., to the right), one
can approximately derive the equation
\begin{equation}
\label{kdv} \frac {\partial u}{\partial \tau} + u \frac {\partial
u}{\partial \xi} + \delta^2 \frac{\partial ^3 u}{\partial \xi^3} = 0
\end{equation}
which is known as the Korteweg-de Vries (KdV) equation \cite{KDV}.
Here $u=\partial x/ \partial \xi, \, \xi=z-c_0t, \, \tau=\bar
\epsilon^\star t, \, \epsilon ^\star = \frac{1}{2} \bar \epsilon
c_0$, and the velocity $c_0$ in our units is 1. The parameter
$\delta^2=a/24\bar\beta$ is the dispersion which plays an essential
role in discrete lattices.

Apart from shallow water waves, the KdV equation is used to describe
hydromagnetic waves in cold plasmas, ion-acoustic waves and long
waves in anharmonic crystals (for references see Ref.\cite{T75}).
Numerical study \cite{ZK65} of this equation in 1965 (see, also,
Ref.\cite{Z68}) has led to the discovery of solitary waves (or
{solitons}), nonlinear waves that propagate through the media
without changing their form. Specifically, it was observed that
starting with the simplest initial condition [$x(z,0)=C\sin \pi z$
with $\dot x(z,0)=0$], solitons appear and strongly interact with
each other, however, after interaction they preserve their
identities. This remarkable property of stability of solitons (see,
also, Ref.\cite{Z67}) was treated as an indication of the existence
of a large number of integrals of motion. Due to the direct
relevance of the KdV to the FPU model, the authors of
Ref.\cite{ZK65} proposed that the recurrent dynamics in FPU lattices
may be explained in terms of solitons as well.

The numerical discovery of solitons \cite{ZK65} has attracted much
attention to the KdV equation and triggered extensive analytical
studies. In particular, in Ref.\cite{L68} it was rigorously shown
that two solitons keep their shape after interacting, and specific
methods were proposed to prove the stability for the general case of
any large finite number of solitons. Further heuristic methods were
developed in Ref.\cite{BK66} to predict the number and speed of
solitons emerging from arbitrary initial conditions. Later, a
nonlinear transformation between the KdV equation and another
nonlinear equation [namely, by changing the term $uu_z$ by $u^2u_z$
in Eq.(\ref{kdv})] was found in Ref.\cite{M68}. After this, a
rigorous method for solving the KdV equation was developed in
Refs.\cite{GGKM67,GGKM68}, by reducing the original nonlinear
problem to a linear one. As a result, complete integrability of the
KdV equation was proved for fixed boundary conditions. Similar
properties of the KdV equation with periodic boundary conditions
have been found numerically in direct numerical simulations
\cite{Z68}.

This remarkable integrability of the KdV equation was used to
propose that similar properties may occur in nonlinear lattices.
Thus, in Ref.\cite{T67} a nonlinear model (Toda-lattice) was
introduced (see, also, Ref.\cite{toda}) with nearest neighbors
interacting through the following potential:
\begin{equation}
\label{toda} U(z)=\frac {a}{b} e^{-bz} + a z ,
\end{equation}
where $a$ and $b$ satisfy $ab > 0$. The corresponding equations of
motion have the form,
\begin{equation}
\label{toda-eq} \ddot x_n = a \left [ e^ {{-b(x_n-x_{n-1})}} - e^
{{-b(x_{n+1} - x_n)}} \right ].
\end{equation}
Formally, this lattice reduces to a harmonic lattice with the spring
constant $\kappa=ab$ in the limit $b\rightarrow 0$, keeping $ab=
const$. One can also see that this model corresponds to the
$\alpha-$model (\ref{alpha}) when $\alpha=-b/2$. Moreover, in the
limit as $b \rightarrow \infty$, the Toda lattice reduces to a hard
sphere systems. Therefore, the model (\ref{toda-eq}) covers two
extreme limits of the interaction, from harmonic to hard-sphere.

The first indication of the integrability of the Toda lattice
appeared in numerical data of Ref.\cite{FST73}. It was shown that
for $N=3$ the trajectories cross the Poincar\'e section in a way
that the corresponding points lie on smooth curves. No evidence was
found of regions with scattered points that is typical of integrable
systems. Moreover, when studying the divergence of neighboring phase
trajectories, linear separations of the trajectories were always
observed. This fact is indicative of the stability of motion, unlike
the opposite case of unstable motion for which the separation of
trajectories increases exponentially with time. Similar behavior was
found for $N=6$ particles. Later, it was rigorously proved
\cite{H74} that this lattice with periodic boundary conditions has
$N$ integrals of motion. Even the initial value problem can be
solved for an infinite Toda-lattice by using the inverse scattering
method, if the motion is restricted to a finite region of phase
space; see details in Ref.\cite{T75}. It is important to stress that
the integrability of motion in the Toda lattice does not prevent
energy sharing among the linear modes (defined in the absence of
nonlinearity). As was shown in Ref.\cite{FST73}, energy sharing
increases with an increase of nonlinearity. Therefore, good
statistical properties can appear in completely integrable systems,
provided the number of degrees of freedom is large, but not for all
quantities.

As a result of the very impressive discovery of integrability of the
KdV and Toda lattices, it was often assumed (and this opinion still
persists in some publications) that the FPU paradox was fully
resolved by the concepts of integrability and solitons. However, the
reality turned out to be even more exciting  because of the direct
relevance of the FPU problem to the dynamical chaos.

\section{V. Strong Chaos}

\subsection{Analytical treatment}

Another approach to the FPU problem is based on the concept of
dynamical chaos. For about ten years after 1955, an understanding of
the fact that dynamical systems with few degrees of freedom may
manifest quite strong irregular motion, turned into systematic
studies of ``stochasticity." This term was used to relate the
irregular motion of completely deterministic systems to that known
for physical systems which are governed by stochastic forces.
Currently, two other terms are widely accepted, {\it dynamical
chaos}, and {\it deterministic chaos}. These terms more correctly
emphasize the purely deterministic nature of chaos. This is in
contrast to conventional statistical mechanics which assumes, {\it
ad hoc}, a probabilistic description of systems due to their
underlying ``randomness". For a long time, the problem of
establishing the conditions under which statistical mechanics is
valid, was one of the central problems in theoretical physics. The
concept of dynamical chaos solves this problem, a fact which is
still not well accepted by physicists although, is quite familiar to
mathematicians. According to the modern viewpoint, classical
statistical mechanics can be considered as a {\it particular case}
of classical mechanics which describes {\it both} regular and
chaotic dynamics. Thus, a statistical description, being useful and
important, is an approximate one, and can be deduced from dynamical
equations of motion under the conditions of dynamical chaos.

One of the important studies of the problem of foundation of
classical statistical mechanics is the work by Krylov \cite{K50}. In
his book, he analyzed the mechanism responsible for statistical
behavior of dynamical systems, which is the exponential instability.
By this term one means that the separation $\Delta (t)$ between two
neighboring trajectories (in phase space) for generic initial
conditions increases in time exponentially, $\Delta (t) \sim \Delta
(0)\exp(ht)$. Here, the rate of the instability, $h$, is called the
``dynamical entropy". Later, this quantity was rigorously studied by
Kolmogorov and Sinai and it then assumed the name
``Krylov-Kolmogorov-Sinai entropy" ({\it KS-entropy}). The
positiveness of this quantity, $h > 0$, currently is used as the
definition of dynamical chaos. Due to this instability, the dynamics
of a system is extremely sensitive to its initial conditions, thus
leading to mixing and other statistical properties (for details see,
for example, Refs.\cite{C79,LL83}).

Numerical and analytical studies of dynamical chaos were strongly
influenced by accelerator physics. As is known, the motion of
charged particles in circular accelerators is affected by forces due
to external magnetic fields that are required for the focusing of
particles in a stable orbit. On the other hand, since the particles
perform many revolutions ($10^{7} - 10^{12}$) around a ring,
nonlinear forces, although weak, are important for a long-term
stability of particle motion. Early studies of nonlinear resonances
in accelerators in 1956-1959 have led to the understanding that they
can result in a kind of irregularity of motion. To the best of our
knowledge, the first observation of this effect refers to the report
\cite{SS56}, in which the authors numerically studied the motion of
electrons in a periodic electromagnetic field. Similar problems of
stability have emerged when studying the motion of electrons in
magnetic traps (see the references and discussions in
Refs.\cite{C69,C79}).

The analysis of the stability of motion of particles in the presence
of nonlinear perturbations has led Chirikov in 1959 to the concept
of the {\it overlap of nonlinear resonances} \cite{C59}. This term
refers to the situation when nonlinear resonances strongly interact
with each other. Specifically, it was found that when the
nonlinearity is weak, one can consider any particular resonance
separately, making use of perturbation theories. However for a
strong nonlinearity, the resonances cannot be treated separately
because in the frequency space (or correspondingly, in action space)
they are very close to each other. Thus, the overlap of resonances
gives rise to the onset of a specific instability which leads to
irregular (chaotic) motion. The analytic estimate for this overlap,
known as the Chirikov criterion, turned out to be very effective for
determining the conditions under which the dynamical chaos occurs in
nonlinear systems \cite{C79}. Results of the first experimental
studies of the nonlinear resonances, their interaction and the onset
of stochastic motion in electron-positron storage rings were
reported in Refs.\cite{KMPT68,KMS69}.

The application of the overlap criterion to the FPU model
(\ref{beta}) has been reported in Ref.\cite{IC65}. According to
Eq.(\ref{betaq}), there are two kinds of nonlinear terms. The term
with $i=j=l$ on the right-hand side plays a specific role and can be
written separately,
\begin{equation}
\label{corr} \ddot Q_k + \omega_k^2 Q_k \left [ 1 - \frac
{3\beta}{4N} \omega_k^2 \left ( 2-\omega_k^2 \right ) Q_k^2 \right]
=\frac{\beta}{8N} \sum_m F_{km} cos \theta_{km} , \,\,\,\,\, \dot
\theta_{km} =\bar\omega_{km}.
\end{equation}
Here $\bar\omega_{km}$ are the exact frequencies (including
perturbation terms) of oscillations of normal modes that slowly
depend on time. One can see that the selected term determines the
nonlinear correction
\begin{equation}
\label{delta_omega} (\delta \omega)_k =  - \frac {3\beta}{8N}
\omega_k^2 \left ( 2-\omega_k^2 \right ) < Q_k^2 >
\end{equation}
to the linear frequency $\omega_k$. Even when small, this correction
cannot be neglected since it depends on the energy of the $k-$th
normal model. Note, that $(\delta \omega)_k $ can be obtained in
first order perturbation theory, by averaging $<Q^2(t)>$ over the
period of the unperturbed motion.

The equations (\ref{corr}) describe the motion of nonlinear
oscillators under the influence of external forces with amplitudes
$\beta F_{km}/8N$. The spectrum of the perturbation due to these
forces is given by the resonance frequencies $\omega_{km}=2\sin
\frac{\pi (k+2m)}{2N}$ with integers $\pm m = 0,1,2,...$
\cite{IC65}. For small values of $k \ll N$ (low modes corresponding
to acoustic waves) the separation between resonances in frequency
space is $ \omega_{k+1}-\omega_k \approx 2\pi/N$ . Therefore, if
nonlinear oscillations of a particular normal mode such as the
maximal shift of frequency is much less than $\Delta_{\omega}$, then
one can neglect the influence of neighboring resonances. In this
case one can obtain the width, $\Delta \Omega$, of a nonlinear
resonance by keeping one resonance term only. The resonance
criterion states that if $\Delta \Omega$ is of the order or larger
than the separation
\begin{equation}
\label{dist} \Delta_{\omega} =\omega_{k+1}-\omega_k
\end{equation}
between neighboring resonances, then trajectories can no longer be
associated with a particular $k-$resonance and can wander between
these two resonances. This transition from one to another resonance
occurs in a quite irregular way, thus resulting in a kind of
diffusion in frequency (or in action) space. In this case strong
energy sharing between resonances is expected.

For acoustic waves with $\Delta_ {\omega} \approx 2\pi/N$ the
critical perturbation for the resonance overlap is defined as
\cite{IC65},
\begin{equation}
\label{crit1}  3\beta_{cr} \frac{E}{N} \sim 3 \frac{ \sqrt {\Delta
k} }{k}
\end{equation}
where $E$ is the total energy of the lattice, therefore, $E/N$ is
the energy per normal mode, and $\Delta k$ is the number of
initially excited modes around the central $k-$mode. Here, units in
which the distance between particles is fixed, $a=1$, are used,
therefore, the length of the chain is $L=N$.

For the case in which high (optical) modes are initially excited,
$N-k \ll N$, the critical perturbation is given by the estimate
\cite{IC65},
\begin{equation}
\label{crit2} 3\beta_{cr} \frac{E}{N} \sim \frac{3\pi^2 \Delta
k}{N^2} \left (\frac{k}{N} \right )^2.
\end{equation}
Note that in this case the mean frequency spacing between nearest
resonances is much less than that for low normal modes, $\Delta
_{\omega} \approx \pi^2/2N^2$, see Eq.(\ref{dist}). For both
acoustic and optical cases, the quantity $3\beta_{cr} \frac{E}{N}
\approx 3 \beta_{cr} \left ( \frac{
\partial x}{\partial z} \right )^2_m $
is the nonlinear term in the corresponding continuous model, which
serves as a control parameter of perturbation.

The above estimates determine the ``stochasticity border" for the
$\beta-$model. From the condition (\ref{crit1}) one can see that for
the lowest value $k=1$ (the most studied case in the FPU model
\cite{FPU55}), one must have a very strong perturbation in order to
observe the non-recurrent behavior. Indeed, the nonlinear term in
energy units in numerical studies of FPU has never exceeded 10\%,
therefore, the system was well below the stochasticity border. This
explains the FPU paradox. On the other hand, from the estimate
(\ref{crit2}) for high modes with $k$ close to $N$, the critical
value of the nonlinear parameter $\beta$ is relatively small and one
can easily observe irregular motion with strong energy sharing among
a large number of high modes. In fact, in a few runs of FPU with
higher modes, one can see a more complex dynamics, and according to
the expression (\ref{crit2}), the parameters used in
Ref.\cite{FPU55} approximately correspond to the border of
stochasticity. It was stressed in Ref.\cite{IC65} that the border of
stochasticity between quasi-periodic and stochastic motion is not
sharp. Rather, it is relatively wide and has a very complicated
structure. For this reason, a strong dependence on initial
conditions is expected in the transition zone.

The analytical estimates obtained above refer to the overlap of two
nearest resonances with $k$ and $k+1$. Therefore, they give the
conditions for the onset of {\it local} stochasticity, and should
not be treated as the threshold of widespread sharing of energy
between {\it all} linear modes. The analysis \cite{IC65} shows that
when only a particular $k-$mode is excited, the stochastic exchange
of energy occurs to higher modes as well, at least for some group of
initial linear modes. Indeed, with the flow of excitation in
$k-$space from low to higher $k$, the border of stochasticity
decreases with increasing $k$, see Eq.(\ref{crit1}). It is
interesting to note that, unlike the case of acoustic waves with $k
\ll N$, if high optic modes are initially excited, strong sharing
between acoustic and optical modes occurs provided that low modes
initially have a small amount of energy.

Based on the resonance overlap approach, a specific study has been
performed in Ref.\cite{I66} for the $\alpha-$model (\ref{alpha}).
The analytical treatment has shown that this model appears to be
much more stable than the $\beta-$model. This is due to the fact
that the nonlinear correction for linear frequencies in the first
order of perturbation theory vanishes for the $\alpha$-model [see
Eq.(\ref{alphaq})] and one needs to consider the second order
approximation. Also, it was found that the most favorable conditions
for the onset of stochasticity in this model occur when initially
the modes with $k\sim N/2$ are excited. To compare with, both the
limits of $k \ll N$ and $k \sim N$ turn out to be more stable. In
general, the analysis of Ref.\cite{I66} indicated that the
$\alpha-$model may be quite close to being integrable and further
numerical studies have confirmed this.

\subsection{Numerical data}

Extensive numerical studies of the $\alpha-$model have been
performed in Refs.\cite{IKC68,IT70,CIT73} exploring the above
analytical predictions. As was discussed in Ref.\cite{IC65}, the
most important property of stochastic motion is the exponential
instability of trajectories with respect to a small change of
initial conditions. To measure this instability, it was proposed to
use the existence of the additional integral of motion in the
FPU-model, namely, parity. As is clear from equations of motion, for
fixed boundary conditions, $x_0=x_{N+1}=0$, there is no interaction
between the even, $k=2, 4, 6, ...$, and the odd, $k=1, 3, 5, ... $,
modes. Therefore, when only odd modes are initially excited, the
energy of the even modes has to be zero. However, in numerical
experiments \cite{IKC68} it was unexpectedly observed that when
exciting the first mode with $k=1$, the energy of even modes is not
exactly zero, and moreover, this energy increases with time. Above
the border of stochasticity the rate of this increase was found to
be exponential, until the energy of the even modes approaches the
energy of the odd modes. The mechanism of this phenomena is due to
round-off errors (of the order of $10^{-19}$) which cannot be
avoided in numerical studies. One should note that these errors are
not random since they are determined by a particular fixed
algorithm. Therefore, they should be treated as a kind of dynamical
perturbation which is not included in the original equations of
motion. As a result, the energy of the even modes can be considered
as a distance (in energy space) between two close trajectories. This
concept  was found to be very useful for determining the degree of
instability.

Given, after some initial time, a small amount of energy in the even
modes ($\sim 10^{-14}$ of the total energy), the rate of instability
for the $\beta-$model (\ref{beta}) has been numerically computed as
a function of the model parameters. Three regions of initial
conditions have been examined: small modes with $k=1$ or $k \ll N$,
high modes with $k \lesssim N$, and the intermediate region with $k
\approx N/2$. In all of these cases quite good correspondence with
the analytical estimates of Ref.\cite{IC65} has been found.
Specifically, for perturbations below the critical value, the rate
of instability was approximately zero. This was in significant
contrast to perturbations above the border, for which strong
exponential instability was easily observed. In order to be sure
that the numerical method used in determining the border is not an
artifact, an additional check was done for two trajectories
belonging to the same parity. The results were found to be analogous
to those obtained from the energy increase of the even parity modes.

In order to study quasi-periodic oscillations for normal modes
involved in the dynamics, in Ref.\cite{BMP73} the possibility of
describing a recurrence using truncated equations of motion was
analyzed. It was found that for typical FPU conditions the dynamics
of the model can be essentially described by a few equations for the
modes close to the initially excited one. Namely, when exciting the
mode with $k=15$ for $N=32$, three (with $k=14-16$) and five ( with
$k=13-17$) coupled equations have been examined numerically.

The important question studied numerically in Ref.\cite{IKC68} is
the dependence of the rate $h$ of instability on the perturbation
parameter $\beta$. As expected from the predictions of
Ref.\cite{C66}, for large values of $\beta$ the dependence
\begin{equation}
\label{exp} h \approx \Delta_{ \omega} \ln
{\frac{\beta}{\beta_{cr}}}
\end{equation}
has been found to correspond to numerical data, with $\Delta_{
\omega} $ as the mean distance between the unperturbed frequencies,
see Eq.(\ref{dist}). On the other hand, when the perturbation was
not very strong, the dependence turned out to be very different and
it can be fitted as $h \approx \Delta_{ \omega}
(\beta/\beta_{cr})^{4/3}$. It was proposed that for a weak enough
perturbation, the instability, being exponential, is due to high
order resonances. Although these resonances are more dense, the
diffusion among these resonance is much slower. Therefore, apart
from the strong stochasticity (chaos) determined by the overlap of
main resonances (due to the first order of perturbation theory), one
can speak about {\it weak chaos} which can also lead (on much larger
time scales) to strong energy sharing between normal modes (see
Sect. VII).

In addition to the instability of motion, in Ref.\cite{IKC68} other
statistical characteristics of the dynamics have been studied as
well: energy sharing among modes, the time dependence of the
energies of each mode, time correlations $<x_n(t) x_n(t+\tau)> $ and
$<E_k(t) E_k(t+\tau)>$ for displacements and energies, as well as
correlations between energies of different modes. The results
strongly support the onset of strong chaos above the border as
analytically predicted in Ref.\cite{IC65}.

An additional numerical study has been reported in Ref.\cite{IT70}
(see, also, Ref.\cite{CIT73}) with higher accuracy and a larger
number of particles (up to $N=500$). These new data confirmed the
main findings of Ref.\cite{IKC68} concerning the border of
stochasticity. Moreover, the exponential dependence (\ref{exp}) for
the rate of exponential instability was also supported by these
data. One of the new observations was the existence of an initial
time scale on which a non-chaotic excitation of modes different from
the initially excited one occurs. After this initial time, a
stochastic exchange of energy begins. Therefore, it was proposed to
modify expression (\ref{crit1}) for the stochasticity border by
using larger values of $k$ due to this effect. This effect seems to
be relevant to the emergence of solitons that occurs in the Toda
lattice. Therefore, the following picture seems to be more correct:
first, an initial regular dynamics occurs in the model, with the
excitation of higher modes. After some initial period of time, a
stochastic exchange between the modes comes into play, with a
practical irreversibility of motion and onset of thermalization.
Note that this thermalization can be restricted to a finite number
of modes, much less than the total number $N$ of degrees of freedom.

To compare with the KdV equation (\ref{kdv}), one should recall that
this equation is an approximation to the FPU model. The main
difference lies in the assumption that the waves traveling in
different directions can be considered independently. This fact may
be crucial in the analysis of the application of the KdV to the FPU
model. In order to check how important the above approximation is,
in Refs.\cite{IT70,CIT73} a specific study of the FPU model with
periodic boundary conditions $x_0=x_N$ was carried out. The quantity
of interest was the emergence of waves traveling in a direction
opposite to that of the initial wave. Specifically, the percentage
of energy of standing waves in comparison with the total energy for
the $k-$th mode was calculated as a function of the perturbation
$\beta$ at some (large) fixed time. It was found that for small
perturbations, opposite moving waves practically do not appear.
However, for $\beta \sim \beta_{cr}$ the amount of energy in the
waves in the opposite direction is of the order of total energy.
This fact demonstrates that for large perturbations the FPU model is
very different from the KdV model.

\section{VI. Further results}

\subsection{Energy sharing and equipartition}

The existence of an initial period of time (``induction period") for
which the motion does not reveal strong energy sharing, was studied
in detail in Ref.{\cite{OHS69} for the $\beta-$model with zero
boundary conditions [see, also, the study \cite{HS69} of the 2D
model]. After this period, strong energy sharing between a large
number of modes was clearly seen, thus corresponding to the
predictions of Refs.\cite{IC65} of the onset of strong
stochasticity. It was also found that this period increases as the
nonlinear coupling decreases. As the criterion for the establishment
of thermal equilibrium, in Ref.\cite{OHS69} velocity-velocity
correlations between close in the chain particles were used. It was
shown that below a critical value of the nonlinear coupling these
correlations are very small; this was used as an indication of
thermal equilibrium. These results were compared with those obtained
for the model with linear coupling only \cite{SH67} for which the
correlations were found to be stronger due to the absence of
ergodicity.

However, the approach of Ref.\cite{OHS69} based on the examination
of correlation functions has been criticized in Ref.\cite{LPRo83}
(see, also, Ref.\cite{BT83}). It was argued that this method does
not give global information about the phase space of the system, and
strongly depends on the choice of correlation functions. Analyzing
other methods for determining the chaotic transition, the authors of
Ref.\cite{LPRo83} proposed to study the distribution of energy modes
after a relatively short period of time. Their analytical analysis
showed that at short times this distribution has an exponential
dependence, $W(k,t) \sim \exp [-B(t)k]$, on the function $B(t)$ that
depends on the model parameters. Numerical data have shown that with
a high accuracy the distribution of energies corresponds to the
analytical expression for $B(t)$. At later times for large enough
nonlinearity the distribution $W(k,t)$ was found to be of the
expected form $W(k,t) \sim 1/k^2$, corresponding to the Boltzmann
distribution of mode energies. These results also confirm the
existence of the stochasticity threshold in its dependence on the
nonlinearity parameter $\beta$.

An interesting quantity to measure the energy sharing has been
proposed in Ref.\cite{LPRo85}. This quantity was found to be quite
useful in the study of relaxation properties of nonlinear lattices.
In order to characterize the energy spread in the mode
representation, the spectral (Shannon) entropy $S(t)$ is used,
\begin{equation}
\label{shannon} S(t)=-\sum_{k=1}^N w_k(t) \ln w_k(t),
\end{equation}
where $w_k=E_k/\sum_i E_i$ is the normalized energy of a particular
mode. The spectral energy is zero when only one normal mode is
excited, and reaches its maximal value $S_{max}=\ln (N/2)$ for
complete equipartition of the total energy among all modes. To avoid
the clear dependence on the number of oscillators, the normalized
quantity
\begin{equation}
\label{eta} \eta = \frac {S_{max} - S_{\infty}}{S_{max} - S(0)}
\end{equation}
was introduced. Here $S_{\infty}$ is the maximum value reached by
$S(t)$ in the time evolution, associated with the ``asymptotic"
value of $S(t)$. This quantity $\eta$ is bounded between zero (which
corresponds to complete ``localization" in one mode), and one (which
corresponds to perfect equipartition).

Computing the normalized spectral entropy $\eta$, in
Refs.\cite{LPRo85,LPRV85} strong evidence in favor of the existence
of an equipartition threshold was given. In numerical simulations,
periodic conditions were used for the $\beta-$model, with initial
excitation of a group of modes $ \bar k \pm \Delta \bar k/2$ with
small values of $\bar k \ll N$. The integration period was chosen
large enough to ensure a practical independence of $\eta$ on time.
These numerical data revealed the remarkable result of a universal
form of dependence of $\eta$ on the energy density $\epsilon = E/N$
for many values of $N$ from $64$ to $512$.  It is interesting to
note that this effect is insensitive to randomization of the FPU
model, for which linear forces were assumed to be different for
different particles (see details in Ref.\cite{LPRo85}). According to
these results, the threshold of equipartition does not disappear in
the large $N-$limit.

Since in numerical computations \cite{LPRo85,LPRV85} the mean value
of $\bar k$ for initially excited modes was taken to be proportional
to $N$ (as well as $\Delta \bar k \sim N$ ), it was claimed, that
the obtained results  are in formal contradiction to the condition
(\ref{crit1}) of a strong chaos, where the threshold disappears with
$\Delta k \sim k \sim N \rightarrow \infty $. The explanation of
this contradiction lies in understanding the meaning of the
stochasticity threshold (\ref{crit1}). Indeed, according to its
derivation, this condition refers to the overlap of nearest
resonances only, and there is no direct relation to energy sharing
among all modes. Although the numerical data show that a large
number of modes appears to share their energy above this threshold,
the question about complete equipartition remains open.

An important comparison of the FPU model with the Toda lattice is
given in Ref.\cite{ILRV86}. As was already discussed (see, e.g.,
Refs.\cite{SOAH70,FST73}), quite strong energy sharing may be
observed in the FPU model in the regime of strong recurrence, below
the border of strong stochasticity. For this reason, the dynamics of
the Toda lattice and FPU models were analyzed from the viewpoint of
equipartition. As was pointed out, one should distinguish between
``energy sharing" and ``equipartition". It was shown that strong
energy sharing can be observed in both models. However,
equipartition occurs in the FPU model only, which is understood to
be non-integrable. The numerical data which demonstrate this
difference are based on the form of the normalized spectral entropy
$\eta$ for large times. Namely, for the Toda lattice the spectral
entropy $S(t)$ never reaches its maximum value, in contrast to the
FPU model for which it does reach the maximum. Therefore, for the
Toda-lattice the dependence of $\eta$ on the energy density
$\epsilon=E/N$ does not show a transition to zero.

\subsection{Stability conditions}

In order to characterize the difference between integrable and
non-integrable lattices, in Ref.\cite{ILRV86} it was proposed to
study these models from the viewpoint of stability. Specifically, it
was observed that the time dependence of the trajectory in the
artificial phase space $\dot \eta(t), \, \eta(t)$ is clearly
different for these two cases. Were the FPU trajectories to appear
irregular and unstable to an external perturbation, for the Toda
lattice the trajectories reveal clear quasi-periodicity and
stability to this perturbation. Thus, in order to distinguish
between integrable and non-integrable chains, a study of the
stability of motion is needed.

The first analytical study of the stability of motion in the FPU
model is reported in Ref.\cite{BMP73}. It was found that the
recurrence dynamics in the $\beta-$model with fixed boundary
conditions may be correctly described by keeping a small number of
equations for those modes that are essentially involved in the
dynamics. For these equations, one can write the condition of a
linear stability which stems from the corresponding Mathieu
equation. According to this condition, a critical value of
perturbation exists above which the motion is unstable. However, the
relevance of this instability to the overlap of nonlinear resonances
remains unclear.

Another approach to instability of motion has been developed in
Ref.\cite{DO76} in application to the KdV equation. It was
analytically observed that certain KdV solutions are unstable. Using
this fact, an attempt was made to relate this instability to that
observed in the FPU lattice. The analytical predictions obtained for
the KdV model have been claimed to explain the instability of motion
in the FPU model. In this study specific initial conditions in the
form of a cnoidal wave were used, for which numerical data
manifested a good correspondence to analytical predictions.

The above analytical studies have suffered from the absence of
reliable expressions that would predict the dependence on the
nonlinear parameter and the number of particles. The first attempt
to shed light to this problem was made in Ref.\cite{BB83} where the
stability condition was obtained for the specific case of the
highest linear frequency, which is initially excited. Note that for
periodic boundary conditions the linear spectrum is doubly
degenerate. Therefore, the highest frequency corresponds to the
middle of the spectrum, $k=N/2$. Using a variational equation for
this mode, a simple approximate formula for the $\beta-$model (in
the large $N-$limit) was derived,
\begin{equation}
\label{bount1} 3 \beta_s \frac {E}{N} \approx \frac {9.7}{N^2}.
\end{equation}
Applying this estimate in Ref.\cite{OHS69} to numerical data
obtained for $N=15$ and zero boundary conditions, some discrepancy
was found. In this respect the authors claimed that their critical
value is, in essence, an upper estimate for the threshold of the
chaotic transition, and cannot be compared with the condition of
widespread energy sharing. Another uncertainty is due to the
different boundary conditions used in the analytical evaluation.
Later, the stability condition (\ref{bount1}) was obtained
\cite{F96} (with almost the same constant) in the more general
context of bifurcations of periodic orbits in nonlinear Hamiltonian
lattices, and with some relation to symmetry breaking and
vibrational localization (breathers).

A similar analysis has been done \cite{BB83} for the $\alpha-$model.
The corresponding stability condition turned out to have the same
form,
\begin{equation}
\label{bount2} \alpha_s \frac {E}{N} \approx \frac {0.84}{N^2}.
\end{equation}
This result is quite unexpected since the $\alpha-$model is assumed
to be closer to being integrable than the $\beta-$model. An
additional study of the stability of the $\alpha$ and $\beta-$models
with attractive potentials, $\alpha < 0$ and $\beta < 0$ was
performed in Ref.\cite{BB83}.

The important results are reported in Ref.\cite{BK84}. Using the
so-called {\it narrow packet approximation}, the following stability
condition was derived for the $\beta-$model:
\begin{equation}
\label{gena} 3 \beta_g \frac {E}{N} \approx \frac {\pi^2}{N^2}.
\end{equation}
It was found that for $\beta > \beta_g$ a parametric instability of
motion emerges for wave packets that populate a number of linear
modes for $k$ within the interval $|k-k_0| = \Delta k \ll k_0$ in a
region of the optical phonon spectrum around $k_0 \approx N/2$. Note
that for the periodic boundary conditions used in Ref.\cite{BK84},
the value $k_0=N/2$ corresponds to the mode with highest frequency,
due to the double degeneracy of unperturbed frequencies $\omega_k$.
The point is that for such initial conditions the FPU model reduces
to the nonlinear Schr\"odinger equation, which is known to be
completely integrable (see details and discussion following). This
fact establishes a link between nonlinear lattices of the FPU type
and models which are now widely used in application to the
Bose-Einstein condensation.

According to the condition (\ref{gena}), if the amplitude of the
initially excited modes exceeds some critical value, the parametric
instability results in a rapid spread of the wave packet over many
linear modes. Even though this packet spreads rapidly, the narrow
packet approximation remains valid for some time. One can see that
formally Eq.(\ref{gena}) coincides with the condition for the onset
of stochasticity due to the overlap of two close nonlinear
resonances, see Eq.(\ref{crit2}). It should be stressed that,
strictly speaking, both conditions correspond to the low boundary
for the emergence of chaos and may be different from those for
strong energy sharing among all modes. Note also that the result of
Ref.\cite{BK84} practically coincides with Eq.(\ref{bount1}), which
also refers to the instability of the highest mode.

It is also very instructive that the condition of the validity of
the narrow packet approximation obtained in Ref.\cite{BK84} reads as
\begin{equation}
\label{validity} 3 \beta_p \frac {E}{N} \approx \frac {\pi^2}{N}.
\end{equation}
This condition may be treated as the critical value above which
strong equipartition among {\it all} modes arises in the lattice.
The important point is that the additional factor $N$ stands in the
denominator of Eq.(\ref{validity}) in comparison with the stability
condition (\ref{gena}). Therefore, one can propose that the
difference between the critical value of perturbation for {\it
local} (chaotic) energy exchange between nearest modes in the
$k-$space, and {\it global} equipartition of energy in the lattice,
is mainly due to this additional $N-$dependence. Note, however, that
what we discuss here refers to the initial excitation of the high
frequencies only. The problem of stability for small values of $k$
(acoustic waves) seems to be very different.

The problem of stability of solutions in the $\beta-$model with
periodic boundary conditions has been studied in a generalized
approach in Ref.\cite{PR97}. A complete rigorous analysis has been
done for the $\pi-$mode considered in Ref.\cite{BB83}, resulting in
an additional correction term which depends on the number of
particles. It was shown that other exact solutions exist below some
critical value of nonlinear parameter $\beta$. Moreover, the
presence of multi-mode invariant manifolds was shown. It was also
pointed out that the relevance of the instability of these solutions
to widespread stochasticity is not clear, although numerical data
generally manifest such a connection [compare Eq.(\ref{bount1}) and
Eq.(\ref{gena}) with Eq.(\ref{crit2}] where $k \approx N$). The
general case of a nonlinear lattice with both $\alpha$ and $\beta$
terms has been under careful study in Ref.\cite{SP94} where, in
particular, the relevance of the instability of the highest
frequency mode to stable localized solutions ( breathers) has been
discussed (see, also, Refs.\cite{BKR90,KL00}). The role of
periodicity of boundary conditions can be examined with the use of
Birkhoff normal forms, see details in Ref.\cite{R02}.

\subsection{Lyapunov exponents }

As is discussed in Ref.\cite{IC65}, the important quantity that
characterizes dynamical chaos, is the local instability for which
two close trajectories in phase space diverge, in time,
exponentially fast. For this reason, many modern numerical studies
are based on the calculation of the rate of this instability
(KS-entropy). A detailed analytical and numerical analysis of the
method of calculating the KS-entropy in many-dimensional dynamical
systems has been performed in Ref.\cite{BGS76}. The approach
developed by these authors was found to be extremely useful in the
study of dynamical chaos in various physical systems. Extensive
numerical analysis of KS-entropy in application to anharmonic chains
with a Lennard-Jones interaction is reported in Ref.\cite{CDGS76}.
This lattice is known to exhibit the stochasticity transition in a
more clear way, compared with the FPU model (see, e.g.,
Ref.\cite{BSBL70}). These data show that the standard procedure of
computing the dynamical entropy due to the average $h \sim <\ln
|d(t)/d(0)|>$ along the trajectories is quite stable with respect to
different kinds of computational errors, and gives reliable results.
However, it should be noticed that in this method the quantity $h$
is not exactly the Kolmogorov entropy, although it often is close to
it. The exact expression for the KS-entropy is a sum of all positive
Lyapunov exponents (LE) (for details and references, see, e.g.,
Ref.\cite{LL83}). Actually, the above method determines the largest
Lyapunov exponent, and this is sufficient to distinguish between
chaotic motion, $h>0$, and (quasi)-periodic, $h=0$, motion. On the
other hand, it was argued in Ref.\cite{LPRo83} that the largest
Lyapunov exponent may not show a correct picture since it does not
experience different ergodic regions. However, in Ref.\cite{CFVo78}
an opposite conclusion was drawn from numerical data, according to
which different non-connected regions of chaos can also be detected,
by searching the fluctuations of the largest Lyapunov exponent.

Since the largest Lyapunov exponent $\lambda_1 \approx h$ can be
used as a measure of chaoticity in a system, analytical estimates of
$\lambda_1$ and its scaling properties are extremely important. In
Ref.\cite{EW88} it was argued that the spectrum of Lyapunov
exponents for long chains may be well approximated by the Lyapunov
exponents of products of independent random matrices, provided the
energy per mode, $\epsilon$, is sufficiently large. This point has
been used in Ref.\cite{CLP95} where an analytical estimate for
$\lambda_1$ was derived in the large $N-$limit. Specifically, a
Gaussian model with noise was used as an approximation of the
dynamical FPU model, and the analytical results were compared with
numerical calculations. An amazingly good correspondence was found
between analytical predictions and numerical data, and two scaling
dependencies were dicovered, $ \lambda_1 (\epsilon) \sim \epsilon^2$
for $ \epsilon \rightarrow 0$ and $\lambda_1 (\epsilon) \sim
\epsilon^{1/4}$ for $\epsilon \rightarrow \infty$. In comparison
with the previously obtained numerical data in Ref.\cite{PL90}, the
first scaling was confirmed. As for the second one, for large
$\epsilon$, their result $\lambda_1 \sim \epsilon^{2/3}$ of
Ref.\cite{PL90} was different. A later theoretical study in
Ref.\cite{DRT97} confirmed the dependence $\lambda_1 (\epsilon) \sim
\epsilon^{1/4}$ by making use of a different approach. An important
point of the studies in Refs.\cite{PL90,CLP95,DRT97} is that there
is a clear transition from one scaling to another. According to
Ref.\cite{DRT97}, this transition occurs at $E \approx
\pi^2/3N\beta$ which corresponds to the onset of strong chaos, see
(\ref{crit2}). This very important fact confirms previous findings
that below the border of strong chaos, found from the Chirikov
criterion, weak chaos persists. Thus, one may expect energy
equipartition for any weak nonlinearity, however, after much longer
times (see discussion below).

Much more information can be drawn from the knowledge of {\it all}
Lyapunov exponents, not only from the largest one. A numerical
method of computing all LE in many-dimensional systems has been
developed in Ref.\cite{BGGS78a,BGGS78b}. Currently, this method is
widely used in many applications, both classical and quantum. Of
particular interest is the distribution of LE as a function of the
index $j$ according to which the Lyapunov exponents are ordered in
an increasing way. Spectrum of the LE has been numerically studied
for the $\beta-$model in Ref.\cite{LPR86} where it was found that
already for $40$ to $60$ particles the limiting distribution
emerges. Thus, these results demonstrate the existence of a
thermodynamical limit for the spectrum of LE. The obtained
distribution was discussed in Ref.\cite{LPR86} with a view towards
its relevance to random matrix approaches.

A very interesting study was reported in Ref.\cite{CCPC97}. As was
already mentioned above, the $\alpha-$ model seems to be more stable
than the $\beta-$model. For the first time this point was noted in
Ref.\cite{I66} where the resonance overlap criterion was obtained
for the $\alpha-$model. The same conclusion can be drawn from the
analysis of nonlinear differential equations of the KdV types, from
the viewpoint of their integrability \cite{Z73}. In order to
quantify the difference between the integrable Toda lattice and
non-integrable $\alpha-$model, in Ref. \cite{CCPC97} the largest LE,
$\lambda_1$, was computed for both models as a function of time. It
was found that at large time scales which are inversely proportional
to the energy density $\epsilon = E/N$, the time dependence of
$\lambda_1 (t)$ is practically indistinguishable for both models.
However, starting from some critical time, a drastic difference is
clearly seen: for the $\alpha-$model $\lambda_1$ tends to a positive
value, in contrast to the Toda-lattice where $\lambda_1$ continues
to vanish with an increase of time. As a possible explanation of
this remarkable effect, the authors conjectured the coexistence of
tiny chaotic regions and relatively large regions of stable motion
in the phase space of the $\beta-$ model. It was argued that the
trajectory might be trapped for a long time in a relatively large
stable regions. Then after some time, the trajectory ``finds" a way
to leave the stable region and enter a stochastic region. Apart from
this suggestion, there is no satisfactory explanation of the
observed effect. Note that the computation performed in
Ref.\cite{CCPC97} was in exact correspondence with the old
computations of FPU, however, with the highest accuracy and for the
longest time ever achieved. Specifically, the lowest frequency mode
with $k=1$ was initially excited.

The observed effect is interesting from many viewpoints. First, it
again confirms the point that the practical difference between
integrable and non-integrable models may be very small, and this
difference can be detected only on very large time scales. Second,
it gives direct evidence of the existence of the threshold of {\it
weak } chaos, although indirect indications for the $\beta-$model
have been reported before (see, e.g. Ref.\cite{KLR94}). Third, it
shows that the largest Lyapunov exponent $\lambda_1$ seems to be the
only quantity which can give reliable results concerning weak chaos.
By searching other initial conditions, the authors of
Ref.\cite{CCPC97} claim that the existence of the stochasticity
threshold $\epsilon_c$ can be clearly seen by examining $\lambda_1$.

Another important result of Ref.\cite{CCPC97} is the $N-$dependence
of the stochasticity threshold for the case when all modes are
initially excited. With an increasing number of particles, it was
found that the scaling dependence $\epsilon_c \sim 1/N^2$ is well
supported by the numerical data. Thus, in the limit $N \rightarrow
\infty $ the threshold in the FPU model vanishes. So far, there is
no satisfactory explanation of this result. Note that most studies
have been done for the $\beta-$model, and it is questionable whether
there are quantitative properties shared by these two models.

\section{VII. Weak Chaos}

As was shown analytically in Ref.\cite{FL70}, nonlinear lattices of
the FPU type have an important peculiarity. Specifically, the
unperturbed motion is linear, therefore, the KAM theory
\cite{K54,A63,M62} formally cannot be applied. Indeed, one of the
conditions for applicability of the KAM theory is that the
unperturbed frequency of oscillations has to be dependent on energy.
For this reason, there is no rigorous ground to expect that under a
very weak perturbation the motion of the FPU model remains stable.
Indeed, the analysis of Ref.\cite{FL70} has shown that the resonance
overlap can occur for arbitrarily small but nonzero values of the
parameters $\alpha$ and $\beta$. Therefore, stochastic motion does
not disappear in the limit of vanishing perturbation, although the
degree of stochasticity may be very small. This situation, known as
``nearly linear oscillations" arises in many practical applications
and requires specific methods of study \cite{I80}.

Another effect which is important in view of discussion about the
behavior of the FPU models for a weak perturbation, is the influence
of nonlinear resonances that appear in high orders of perturbation
theory. Indeed, the overlap criterion \cite{IC65} is based on
consideration of the first-order resonances only. On the other hand,
second-order resonances may create a wide resonance net and lead to
strong equipartition which can occur at much larger time scales.
This problem was briefly discussed in Refs.\cite{IC65,IKC68}, and
further numerical studies have confirmed the point that apart from
{\it strong} chaos (with a fast equipartition and large values of
the Lyapunov exponents), one can speak about {\it weak chaos}. This
weak chaos is characterized by a much weaker equipartition of
energy, smaller values of LE, and a different kind of scaling of the
LE. Recently, the role of these high-order resonances was examined
in Ref.\cite{S97}, together with a detailed analysis of
peculiarities related to the nearly-linear character of the
oscillations. The $\alpha-$model was examined and the stochasticity
threshold has been found for both strong and weak interactions.

The concept of weak chaos poses an important question of whether the
threshold of stability for an infinite time scale vanishes in the
thermodynamical limit, $N \rightarrow \infty$. To shed  some light
in this problem, in Ref.\cite{K89} numerical simulations have been
made for the $\beta-$model (with periodic boundary conditions)
paying main attention to the long-time behavior. Measuring the
spectral entropy (\ref{eta}), it was found that the result strongly
depends on whether initially one mode, $\Delta k = 1$ or a group of
modes with $\Delta k \gg 1$, is excited (with both $k$ and $\Delta
k$ proportional to $N$). In the first case, the data indicate the
existence of a threshold $\beta E/N \sim const$. In the second case
the threshold vanishes as $\beta E/N \sim 1/N^{\gamma}$ with $\gamma
\approx 1$. As one can see, these results are in contradiction with
the estimate (\ref{crit1}) obtained for resonance overlap. The
existence of finite times of relaxation to the equipartition has
also been confirmed in Ref.\cite{LLR99} where the strong influence
of initial transient times has been detected, after which generic
behavior emerges with no dependence on initial conditions.

One specific question widely discussed in the literature is how fast
the relaxation is to a steady-state distribution of energy among
normal modes. An extensive study of this problem was performed in
Ref.\cite{PL90} (see, also, Ref.\cite{GGMV92}). The authors made an
attempt to relate the data for time dependence of the spectral
entropy $\eta(t)$, to the estimates of Refs.\cite{N77,N79,BGG85} for
the rate of the Arnold diffusion. In 1964 Arnold has proved
\cite{A64} that many-dimensional nonlinear systems are, in general,
globally unstable due to a very peculiar diffusion ({\it Arnold
diffusion}), for details see, e.g. Refs.\cite{C79,LL83}). Loosely
speaking, this diffusion occurs (below the border of resonance
overlap) for initial conditions {\it inside} the narrow stochastic
layers which surround any nonlinear resonance. Due to an everywhere
dense set of resonances (of different orders), starting from a point
inside this {\it Arnold web}, the trajectory diffuses over the web
{\it along} these resonances. Although Arnold diffusion is extremely
weak (exponentially small in the perturbation parameter), the motion
is unbounded in the phase space of the system. There is a widespread
belief (although still there are no reliable numerical results) that
Arnold diffusion is responsible for a weak instability in FPU
lattices. By fitting data to Nekhoroshev's expressions, in
Ref.\cite{PL90} the following empirical dependence has been
obtained:
\begin{equation}
\label{relax1} \eta(t) = exp \left [ -\left (t/\tau \right)^\nu
\right ]
\end{equation}
for $t< \tau_R$, and
\begin{equation}
\label{relax2} \eta(t) = \eta_{\infty} \equiv exp \left [ -\left
(\tau_R/\tau \right)^\nu \right ]
\end{equation}
for $t\geq \tau_R$, with the numerical estimate of $0.3 \leq \nu
\leq 0.5$. As for the relaxation time $\tau_R$, it is argued that it
is proportional to the energy density, $\tau_R \sim \epsilon$, and
therefore, to the number of particles $N$ for fixed total energy.
The nonzero value of $\tau_\infty$ is discussed in Ref.\cite{PC91}.
Among others, the most realistic explanation of this result is that
it is due to fluctuations of the energies $E_k$, which are not taken
into account in the normalization factor for $\eta$. One of the
important conclusions drawn in Refs.\cite{PL90,PC91} is that the
equipartition of energy is always reached. This supports the
expectation of non-existence of a minimal critical value of
nonlinearity for the stochasticity. Another conclusion is that the
critical value of perturbation $\epsilon_c$ which marks the
transition from weak to strong stochasticity, corresponds to the
overlap condition (\ref{crit1}).

The region of weak stochasticity was closely examined in
Ref.\cite{LLL95}. For initial conditions the low modes were excited
in the $\beta-$model with zero boundary conditions. As was shown
numerically, for low initial energy the distribution of mode
energies after large time follows an exponential decrease with
increasing the mode number. This exponential dependence was
explained theoretically, by finding an approximate solution of
equations of motion written in a form similar to that of a nonlinear
Schr\"odinger equation. It was also shown numerically that a single
nonlinearity parameter,
\begin{equation}
\label{licht} R = 6\pi^{-2} \beta E_{\gamma} (N+1),
\end{equation}
governs the local interactions between the low $k\ll N$ modes. For
$R\gg1$ there is a critical energy $E_c \approx 2.8$ (for $\beta
=0.1$) for which strong diffusion in $k-$space leads to
equipartition. Below this border the diffusion leads to an
exponential distribution of mode energies. It is noted that $R$
being sufficiently large corresponds to the overlap criterion of the
onset of widespread stochasticity. In further studies \cite{LLR95}
the energy transitions and different time scales have been
classified and discussed, as a function of energy. These results
have been generalized in Ref.\cite{LLR96} to FPU-type lattices with
two types of masses randomly distributed along the chain.

The dependence on initial conditions is another important question.
One can classify the following cases: (i) single mode excitation
with small $k$; (ii) a group of low frequency modes with $\Delta k$
and $k$ proportional to $N$ (thermodynamic limit); (iii) single mode
excitation with $k \approx N/2$ (narrow packet approximation), and
(iiii) high frequency excitation with $k \approx N$. As one can see,
other important cases are missed in this picture, thus showing how
difficult, in general, the problem is of statistical relaxation in
nonlinear lattices. A comparison of cases (i) and (ii) was done in
Ref.\cite{LLR95}. It was found that transient times to equipartition
in case (ii) are proportional to $\sqrt N$, in comparison with
single mode excitations (i) where this time does not exist or is not
important. The transient times are characterized by non-universal
dynamical characteristics, in contrast with larger times for which
one can find a good scaling dependence on the energy density $E/N$.
Concerning the case (iiii), one can refer to Ref.\cite{ULC00} where
the energy equipartition for initially excited high frequency modes
was studied. As for the case (iii), the important findings are
analytical ones, with much attention given to the instability
conditions.

\section{VIII. NSE and BEC}

As was found in Ref.\cite{BK84}, one of remarkable properties of the
$\beta-$model (\ref{beta}) is its direct relevance to the nonlinear
Schr\"odinger equation (NSE). Indeed, let us write the Hamiltonian
corresponding to the equations of motion (\ref{beta}),
\begin{equation}
\label{Ham0} H = \sum_{n=1}^{N} \left[ \frac {p_n^2}{2} +
\frac{1}{2} \left( x_{n+1} - x_n \right)^2 + \frac{ \beta}{4}
\left(x_{n+1} - x_n \right)^4 \right].
\end{equation}
In the following, we consider periodic boundary conditions,
$x_0=x_N$. In analogy with the quantum mechanics, we use the
canonical variables $a_k$ and $a^{\star}_k$,
\begin{equation}
\label{ak} a_k=\frac{1}{\sqrt{2\omega_k}} \left({\cal P}_k - i
\omega_k {\cal Q}_k^{\star} \right)
\end{equation}
where
\begin{equation}
\label{pq}{\cal P}_k= \frac{1}{\sqrt N} \sum_{n=1}^{N}p_n
e^{-i\frac{2\pi kn}{N}}\,;\,\,\,\,\,\,\,\,\,\,\,\,\,\, {\cal
Q}_k=\frac{1}{\sqrt N} \sum_{n=1}^{N}x_n e^{i\frac{2\pi kn}{N}}
\end{equation}
with $\omega_k=2 \sin {\frac {\pi k}{N}}$ as the frequency of the
$k-$th linear mode, $k=1,2,...,N$. Assuming that the initial packet
in $k-$space is narrow and centered at $k_0\approx N/2$, it can be
shown \cite{BK84} that the Hamiltonian takes the form,
\begin{equation}
\label{H1} H= \sum_{k=1}^N \omega_k a_k^{\star} a_k + \frac{1}{2}
\sum_{k_1 k_2 k_3 k_4}V_{k_1 k_2 k_3 k_4}
a_{k_1}^{\star}a_{k_2}^{\star}a_{k_3}a_{k_4}
\delta_{k_1+k_2-k_3-k_4} + {\cal O}(1).
\end{equation}
Here the term $V_{k_1 k_2 k_3 k_4}=V_0 + W_0(q_1+q_2+q_3+q_4)$ with
$V_0=\frac {3\beta}{N}\left( \sin{\pi \frac {k_0}{N}} \right)^2$ and
$W_0=\frac {3\pi \beta}{4N^2} \sin {\frac {2\pi k_0}{N}} \ll V_0$
describes the ({\it resonant}) four-wave interaction
\cite{SUZ88}(with $q=k-k_0 \ll k_0$). All other ({\it non-resonant})
terms of ${\cal O}(1)$ can be neglected in this approximation. By
expanding $\omega_k$ at the point $k_0$, one can obtain, $\omega_k
\approx \omega_{k_0} + \Lambda q - \Omega q^2$. The parameters
$\Lambda$ and $\Omega$ are $\Lambda = \frac {2\pi}{N} \cos {\frac
{\pi k_0}{N}} \approx \left ( \pi/N \right) ^3$ and $\Omega
=\left(\frac {\pi}{N}\right )^2 \sin {\frac {\pi k_0}{N}} \approx
\left ( \pi/N \right) ^2$. As a result, the equations of motion $i
\dot a_k =
\partial H / \partial a_k^{\star}$ can be written as
\begin{equation}
\label{eqA} i\dot A_q = -\Omega q^2 A_q + V_0
\sum_{q_1,q_2,q_3}A^{\star}_{q_1}A_{q_2}A_{q_3}
\delta_{q+j_1-j_2-j_3}
\end{equation}
where $A_q = \exp \left ( i(\omega_{k_0} + \Lambda q )t \right)
a_{q+k_0}$. As one can see, these equations describe a nonlinear
chain of interacting oscillators. Using the transformation $
\Phi(\theta,t) = \sum_q A_q(t) \exp (iq\theta)= \Phi(\theta+2\pi,
t)$, we obtain the NLS equation,
\begin{equation}
\label{nls} i\frac {\partial \Phi}{\partial t} = \Omega
\frac{\partial^2 \Phi}{\partial \theta^2} +V_0 |\Phi|^2 \Phi.
\end{equation}
This {\it classical} equation is well known in the physics of
interacting particles and widely discussed in many applications. It
is a particular case of the Gross-Pitaevskii (GP) equation
\cite{G61,P61} which attracts much attention in connection with
Bose-Einstein condensation (BEC). Using the mean-field
approximation, this equation describes the evolution of the
condensate wave function, and, in essence, is of a semi-classical
nature. The important peculiarity of the GP-equation is its complete
integrability (see, e.g., Ref.\cite{DEGM82}). This is of special
interest from the point of view of the statistical properties of the
condensate. As was confirmed numerically \cite{BK84}, in the FPU
$\beta-$model good conservation of first three integrals of motion,
analytically derived for the GP-equation, can be observed provided
the initial packet is centered at $k_0=N/2$.

The model of type (\ref{H1}) was recently examined using close
numerical investigations \cite{VL00}. Specifically, the dynamics of
the condensate in one-dimensional geometry was assumed to be
governed by the following Hamiltonian in the action-angle
representation, $A_n=\sqrt{I_n} \exp{(i\theta_n)}$:
\begin{equation}
\label{leven} H= \sum_n \omega_n I_n + \frac {g}{2} \sum_{n_1 n_2
n_3 n_4} V_{n_1 n_2 n_3 n_4}
\left(I_{n_1}I_{n_2}I_{n_3}I_{n_4}\right) ^{1/2} e^{-i\left(
\theta_{n_1} + \theta_{n_2} -\theta_{n_3} -\theta_{n_4} \right)}.
\end{equation}
Here $\hbar \omega_n = n^2 \pi^2 \hbar^2 /2mL^2$ labels the
single-particle energy levels, $m$ is the mass of particles, $L$ is
the length of the one-dimensional box, and $V_{n_1 n_2 n_3 n_4}$
corresponds to the matrix elements $\int_0^L\phi_{n_1}
\phi_{n_2}\phi_{n_3}\phi_{n_4} dz $ of the interaction, with
$\phi_n$ as the normalized modes of the box (see details in
Ref.\cite{VL00}).

In the study of the model (\ref{leven}), methods developed for
classical nonlinear lattices have been extensively used. Of
particular interest was the time-dependence of the normalized
spectral entropy (\ref{eta}) since it can be used to distinguish
between regular and irregular dynamics of the condensate. Note that
due to the zero boundary conditions used in Ref.\cite{VL00}, the
dynamics may be non-integrable in contrast to the GP-equation. For a
weak interaction the dynamics was numerically found to be
quasi-periodic for generic initial conditions, which may be compared
with the recurrent motion in the FPU models. On the other hand, with
increased interaction strength, the dynamics appears to reveal
chaotic properties. This effect was accompanied by a strong decrease
of the spectral entropy $\eta(t)$, and by irreversible dynamics. The
main result is that starting from conditions in which the low modes
were initially excited, the energy diffusively spreads over modes
with higher frequencies. The stochasticity was found to be stronger
for low excited modes, in contrast to the FPU model. This fact may
be explained by the different kinds of the unperturbed frequency
spectrum in these two models.

The above results show that many of properties of classical
nonlinear lattices, especially, non-integrable models of the FPU
type, can be discussed in a more general context, namely, in the
context of the physics of interacting quantum particles. Although
the latter subject is not new, the problem of the onset of {\it
many-body chaos} in quantum systems of interacting Fermi and Bose -
particles has recently attracted much attention due to its important
physical applications (see, e.g. Ref.\cite{QC}). In this respect, an
interesting study was reported in Ref.\cite{BIKT86,BT04} where the
quantum analog of Eq.(\ref{eqA}) was introduced,
\begin{equation}
\label{quantum} i \dot A_j = -j^2 (1+q) \Omega A_j + \hbar V_0
\sum_{j2,j3,j4} A_{j_2}^{\dag} A_{j_3} A_{j_4} \delta_{j+j_2 -j_3
-j_4}
\end{equation}
Here $\left [ A_j, A_k^{\dag} \right ]=\delta_{jk}$ and $q= \hbar
\beta \cot (\pi/2N)/32 N$, with $\omega$ and $V_0$ defined as in the
classical model.

Analytical studies \cite{BIKT86,BT04} of the dynamical instability
of motion gave quite unexpected results. As was already discussed in
Section VI, in the narrow packet approximation the classical
dynamics of the FPU model displays an instability above the critical
value given by Eq.(\ref{gena}). In contrast to this result supported
by other studies, in the quantum model the instability occurs for
any weak perturbation. On the other hand, the rate of this
instability turns out to be slower than for the classical model. If
the latter effect could be explained by quantum suppression of
classical instability (as in the Kicked Rotor model
\cite{CCIF79,CIS81,I90}), the absence of the instability threshold
is a somewhat new effect. One can suggest that this effect is due to
quantum tunneling in the effective potential, and further studies
appear to be very important.

The above approach has been recently developed and applied in
Ref.\cite{BSB02} to the problem of collapse in the Bose-Einstein
condensation with an attractive potential. As is known, the collapse
dynamics cannot be described by the Gross-Pitaevskii equation,
therefore, new ideas are important. As was shown in
Ref.\cite{BSB02}, near the instability threshold quantum effects
turn out to play an important role and have to be taken into
account. Specifically, when comparing the GP and quantum dynamical
growth of the unstable modes, the absence of the instability
threshold was confirmed for the quantum model, in contrast with the
GP equation. This means that the quantum solution is always
unstable, and eventually collapses in a finite time. The difference
between the GP and the quantum model is important when approaching
the classical (GP) critical limit of the instability. This effect
may be compared with an exponentially fast spread of wave packets in
quantum systems which are chaotic in the classical limit. As was
shown in Refs.\cite{BZ78a,BZ78b}, for classically chaotic systems
quantum effects are extremely strong and reveal themselves on a very
short (logarithmic) time scale. Recent rigorous results \cite{BV03}
on the quantum instability of averages near stationary points seem
to have a direct relevance to the more general problem of quantum
corrections in the region of chaos.

It would be interesting to study the influence of terms, neglected
in the narrow packet approximation, when obtaining the NLS equation
(\ref{nls}). It is easy to show that by keeping the next terms in
the expansion of $\omega_k$ and $V_{k_1,k_2,k_3,k_4}$ about the
point $k_0$, one can derive the following equation:
\begin{equation}
\label{newEQ} i\frac {\partial \Phi}{\partial t} = \Omega
\frac{\partial^2 \Phi}{\partial \theta^2} +V_0 |\Phi|^2 \Phi + i
\chi \frac{\partial^3 \Phi}{\partial \theta^3} -4i W_0|\Phi|^2
\frac{\partial \Phi}{\partial \theta},
\end{equation}
which describes the dynamics of the FPU model more correctly than
the Gross-Pitaevskii equation (with $\chi=\frac{1}{6}\partial^3
\omega_k/\partial k^3$ at $k=k_0$). The additional terms in the
above equation may be important for describing the evolution of wave
packets for large time scales, and for the breakdown of
integrability.

One recent attempt to consider the dynamics of the Bose-Einstein
condensate by making use of an approach developed in the field of
{\it quantum chaos}, was performed in Ref.\cite{BBIS04}. Using
numerical simulations, the authors focus on the dynamics of the
Bose-Einstein condensate on a ring, which is described by the
quantum Hamiltonian,
\begin{equation}
\label{Qham} \hat H = \sum_k \epsilon_k \hat n_k + \frac{ g}{2L}
\sum_{k,q,p,r}\hat a_k^{\dag} \hat a_q^{\dag} \hat a_p \hat a_r
\delta_{k+q-p-r}.
\end{equation}
Here $\hat n_k=\hat a_k^{\dag}\hat a_k$ is the occupation number
operator, $\hat a_s ^{\dag}$ and $\hat a_s$ are the
creation-annihilation operators, and $\epsilon_k=4\pi^2k^2/L^2$. As
one can see, this Hamiltonian can be considered to be the quantum
version of the classical Hamiltonian (\ref{H1}) describing the
evolution of the FPU model in the narrow packet approximation. The
behavior of this system is governed by only one parameter $n/g$,
where $n$ is the particle density on a ring of length $L$, and $g$
is the strength of the interaction between bosons, determined by the
interatomic scattering length.

As is known, for weakly interacting particles, $n/g \rightarrow
\infty$, the mean-field approximation gives the correct description
of the dynamics. In the other limit of strongly interacting
particles, known as the Tonks-Girardeau regime, $n/g \rightarrow 0$,
the density of interacting bosons becomes identical to that of
non-interacting fermions. The transition between these two regimes
is known to correspond, approximately, to $n/g \approx 1$.

The main interest in Ref.\cite{BBIS04} was to observe and quantify
the degree of irregularity in the dynamics of the condensate.
Specifically, the situation in which all bosons initially occupy the
single-particle level with the angular momentum $k=0$ has been
explored, with a further analysis of the evolution of the condensate
in time. The main result of the study in Ref.\cite{BBIS04} was that
with an increase of the interaction strength $g$, regular
(quasi-periodic) dynamics alternates with irregular behavior of the
observable quantities. This transition was found numerically to
occur at the transition from the mean-field to the Tonks-Girardeau
regimes. Given the clear evidence of the efficiency of the proposed
approach, these results open the door for further studies of the
condensate dynamics from the viewpoint of many-body chaos.

\section{IX. Concluding remarks}

Due to space limitations, many recent studies are not discussed
here, although they are relevant to the FPU model. In the first
line, one should mention an increasing interest in the existence of
localized nonlinear oscillations (breathers) emerging in nonlinear
lattices (for a review, see, e.g. Ref.\cite{FW98}). For some time,
their existence in the FPU model was questionable, mainly due to the
fact that the main interest initially was related to low-frequency
excitations. As was shown in the early paper \cite{ZD67}, localized
optical excitations (high-frequency modes) can be observed in the
FPU model with alternative masses. This was the first indication
that by exciting the highest modes in the FPU lattice, a new kind of
solution with special structure emerges (the most recent results on
the diatomic FPU model are reported in Ref.\cite{MZT03}). Although
self-localized solitons in anharmonic lattices without impurities
were predicted quite a long time ago in Refs.\cite{ST88,SPS92}, only
recently the existence of breathers in FPU lattices has been proved
rigorously (see, e.g., Refs.\cite{J01}). In connection with the FPU
problem one should mention the results of Refs.\cite{TA96,BVAT99}
where it was shown that breathers can be responsible for the slow
relaxation of initially thermalized nonlinear lattices. This effect
seems to be relevant to the long-term regular dynamics in the FPU
models, which is alternated by a strong energy sharing between
linear modes. To date, many studies of breathers (including {\it
chaotic breathers}, see in Ref.\cite{CDRT98}) in the FPU model have
been carried out, and we hope that the reader can find in this issue
more information on the subject.

Coming back to the original question about the ergodicity and
thermalization in the FPU model, one should conclude that some of
problems still remain open. In particular, the existence of the
threshold for weak chaos in the thermodynamical limit $N \rightarrow
\infty$ is still under study by many researchers. As is clear from
the above discussions, the main difficulty, apart from the numerical
one, is the strong dependence of the results on the model
parameters. The behavior of the model depends strongly on whether
low- or high-frequency modes are initially excited. Also, the number
of excited modes seems to be important for the dynamics, as is
indicated in previous studies. Last, but not least, is the fact of a
difference between the $alpha-$ and $beta-$models. Therefore, future
studies are desirable, both analytical and numerical.

One should mention the direct relevance of the FPU model to the
models of the Bose-Einstein condensation. As was shown in
Ref.\cite{BK84}, the narrow packet approximation in the
$\beta-$model leads to the Gross-Pitaevskii equation with an
attractive potential. Thus, the instability of highest modes in
periodic FPU lattices corresponds to that of the Bose-Einstein
condensate. This fact is important for further studies of
instabilities both in the nonlinear classical lattices and in
quantum models of the Bose-Einstein condensate. Note that now it is
possible to control experimentally the sign of the interaction
between bosons, and to observe the collapse of the condensate (see,
for example, Ref.\cite{Do01}).

Another new direction is the study of dynamics of quantum models of
interacting Bose particles. In particular, direct quantization of
classical nonlinear chains related to the Gross-Pitaevskii equation
shows a close analogy for the dynamics in classical and quantum
models. Recent numerical data \cite{BBIS04} for the transition
between the mean field and Tonks-Girardeau regimes have revealed the
onset of irregular motion of the condensate. This type of transition
is known to occur in quantum models of interacting particles which
are chaotic in the classical limit. Therefore, one can expect that
methods well developed in the theory of quantum chaos, may give new
insight on the dynamics of interacting bosons in the condensates.

The above practical problems are part of the more general problem of
quantum-classical correspondence for systems with irregular
behavior. From this point of view, the FPU model is of particular
interest and can be considered as an important example. As discussed
in Ref.\cite{C97}, there are two mechanisms which are responsible
for the appearance of statistical behavior of dynamical
(deterministic) systems. The first mechanism is the thermodynamic
limit with $N \rightarrow \infty$, which is well known since the
early days of statistical mechanics. The important point here is
that this limit has nothing to do with chaos, it is based on the
ergodicity of motion only, which is known to be the weakest
statistical property. As we already discussed, perfect statistical
and thermodynamical properties are known to emerge even in
completely integrable systems such as the Toda-lattice. Although
there are initial conditions which correspond to solitons, the
measure of these specific conditions is extremely small, and surely
can be neglected practically. Therefore, the role of additional
terms that break the integrability becomes important when the number
$N$ is finite. The expectation of FPU was based on their belief that
$N=32, 64$ is large enough in order to observe the equipartition in
the presence of small nonlinearity.

The second mechanism for the onset of statistical properties in
dynamical systems is, in principle, different. It is based on a
local instability of motion for generic initial conditions in the
phase space of the system. With this mechanism the ergodicity is not
important provided the total measure of initial conditions with
regular motion is very small, although it can be finite. Due to this
local instability (with reflecting boundaries in phase space), this
motion reveals clear mixing properties, leading to strong
sensitivity of the motion to initial conditions. As a result, an
apparent irreversibility of motion emerges since any weak external
perturbation gives rise to non-recurrence of the initial conditions.
The important point for this scenario is that the dynamical chaos
can emerge in systems with few degrees of freedom, in contrast to
the first (thermodynamic) mechanism. It is important to stress that,
although these two mechanisms are different, the common feature is
that in both cases the time dependence of the observables can be
described by an infinite number of statistically independent
frequencies (see details in Ref.\cite{C97}).

Turning to quantum systems, the origin of ``quantum chaos" in
dynamical systems is based on the first mechanism, with no relevance
to the local instability of trajectories. Specifically, irregular
behavior of a system emerges when an initial wave packet consists of
many exact {\it chaotic} eigenstates with statistically independent
frequencies (see, e.g. Ref.\cite{FI97}). This concept is very
important for establishing the conditions for the onset of chaos in
quantum systems and in quantifying their irregular properties.
Therefore, the understanding of physical effects found in the FPU
model, as well as the use of tools developed for identifying these
effects, may be useful for the study of quantum systems with complex
behavior.

As a consequence of their research on the FPU model, in this brief
review the authors have tried to summarize the main ideas, tools and
results related to the FPU paradox, after 50 years of the celebrated
paper. For one of the authors (FMI), the FPU problem was his initial
scientific PhD research. For GPB, the relation discovered between
the FPU model and nonlinear Schr\"odinger equation has contributed
to his interests in Bose-Einstein condensation problems. And for
both of the authors, the preparation of this manuscript provided a
welcome opportunity to learn much more about new trends in the
physics of nonlinear phenomena.

\section{acknowledgments} The authors very much appreciate the
opportunity to have worked with Boris Chirikov in the Institute of
Nuclear Physics in Novosibirsk, Russia. They are also grateful to
A.~R.~Bishop, D.~K.~Campbell, G.~D.~Doolen, R.~E.~Ecke for fruitful
discussions. This work was supported by the Department of Energy
(DOE) under Contract No. W-7405-ENG-36.

\bibliography{basename of .bib file}

\end{document}